# A State Representation Approach for Atomistic Time-Dependent Transport Calculations in Molecular Junctions

Tamar Zelovich[1], Leeor Kronik[2], and Oded Hod[1]

*1) Department of Chemical Physics, School of Chemistry, The Raymond and Beverly Sackler Faculty of Exact Sciences, Tel Aviv University, Tel Aviv 69978, Israel*

*2) Department of Materials and Interfaces, Weizmann Institute of Science, Rehovoth 76100, Israel*

Abstract:

We propose a new method for simulating electron dynamics in open quantum systems out of equilibrium, using a finite atomistic model. The proposed method is motivated by the intuitive and practical nature of the driven Liouville von-Neumann equation approach of Sánchez et al. [J. Chem. Phys., 124, 214708 (2006)]. A key ingredient of our approach is a transformation of the Hamiltonian matrix from an atomistic to a state representation of the molecular junction. This allows us to uniquely define the bias voltage across the system while maintaining a proper thermal electronic distribution within the finite lead models. Furthermore, it allows us to investigate complex molecular junctions, including non-linear setups and multi-lead configurations. A heuristic derivation of our working equation leads to explicit expressions for the damping and driving terms, which serve as appropriate electron sources and sinks that effectively "open" the finite model system. Although the method does not forbid it, in practice we find neither violation of Pauli's exclusion principles nor deviation from density matrix positivity throughout our numerical simulations of various tight-binding model systems. We believe that the new approach offers a practical and physically sound route for performing atomistic time-dependent transport calculations in realistic molecular junctions.

# 1. Introduction

The ability to exploit molecules as miniature electronic components was first theoretically predicted in a seminal paper by Aviram and Ratner.[1] The invention of breakthrough technologies for the visualization and manipulation of systems at the molecular level brought this prediction to realization. Currently, the fabrication of molecular junctions for conductance measurements is routine practice in many research laboratories (see, e.g. Refs. 2-7). Apart from downscaling electronic devices and gaining higher computational efficiency, the use of molecules as electronic components may give rise to novel functionalities based on their quantum mechanical nature, thus redefining electronics.[8-13]

The miniature nature of molecular electronics assemblies allows for a unique interplay between the availability of high resolution experimental data and highly accurate theoretical treatment of key transport phenomena. A major challenge for modeling electronic transport through molecular junctions is the ability to provide an appropriate atomistic out-of-equilibrium description of the entire lead-molecule-lead system. This is often addressed by replacing the full (infinite in principle) system by a finite system with open boundaries. The finite system encompasses the molecular entity and a finite section of the lead models and is often termed the "extended molecule". The challenge now appears in the need to provide an appropriate description of the open boundary conditions of the finite system. One of the most widely used approaches to address this challenge is the simple (yet powerful) steady-state picture of Landauer.[14, 15] Within this approach, the conductance is related to the electron transmittance probability through the molecular junction which, in turn, is often calculated via non-equilibrium Green's function (NEGF) methods.[16, 17] In the latter, the open nature of the system is addressed by the concept of the leads' self-energies, which expresses the true influence of the semi-infinite leads by means of an exact effective interaction term acting on the finite system. Over the past two decades, this approach has proven to be highly successful in describing electronic transport through mesoscopic systems[16, 18, 19] as well as molecular junctions.[9, 20, 21]

The original formulation of the Landauer approach refers to steady-state transport and therefore cannot address issues of electron dynamics, which are of crucial importance when studying transient effects, external time-dependent fields, and coupled electron-nuclei dynamics. However, it can be extended to address dynamical effects by using sophisticated time-dependent Green's function approaches.[22-28] These methods have been applied successfully to relatively simple model systems in order to gain insights regarding the important physical phenomena governing the time-dependent propagation of electrons through narrow molecular bridges.[22, 23, 29] However, such calculations rely on a proper description of the out-of-equilibrium electronic structure of the model system and on a sufficiently accurate calculation of the self-energies of the leads. Both issues are very demanding computationally, thus restricting the practical applicability of these approaches. Alternative numerically exact approaches, based on real-time path-integral Monte-Carlo[30-32] and multilayer multiconfiguration time-dependent Hartree simulations[33, 34] have been developed to treat electron dynamics in open many-body systems. These innovative methods can provide important insights when applied to phenomenological models of molecular electronic junctions. However, they are computationally demanding and can therefore be of limited applicability in realistic molecular junction models.

In light of the difficulties encountered by these formally exact methods, a hierarchy of approaches that aim to mimic the out-of-equilibrium open system by means of a closed one have emerged, offering different points of compromise between accuracy and computational burden. These approaches are not limited to a specific description of the electronic structure of the underlying system and can, in principle, be applied with any Hamiltonian representation of the system, ranging from a simple tight-binding description to the most advanced first principles methods.

For steady state currents, several such approaches that circumvent the explicit treatment of the semi-infinite leads have been suggested. One example is the Lagrange multiplier method where a constant current is enforced on the system.[35-38] The method of source-sink potentials is a second example where the Hamiltonian of the finite system is augmented by source and sink terms that serve to inject and absorb electrons near the physical boundaries of the finite system.[39-43]

One of the earliest time-dependent approaches developed along these lines suggested that the explicit calculation of leads' self-energies can be avoided by approximating them with complex absorbing potentials.[44-46] Here, the time-dependent response of the finite system to an external electric field perturbation[44, 45] or local edge potentials[46] is simulated and the excited electrons are absorbed near the boundaries, thus dissipating the excess energy that they carry and avoiding back-scattering into the system. This method was successfully used to analyze the effects of geometrical constriction on the transport properties of the nanoscale junction. However, it is suitable for short simulation periods, during which no significant depletion of the electron density due to the absorbing boundary conditions occurs.

Several other methods that avoid the explicit treatment of the semi-infinite electronic reservoirs have been developed. One such approach is the momentum space method,[47] where the time-dependent Kohn-Sham equation, augmented by local edge potentials, is Fourier-transformed to momentum space where the boundary conditions on the wave functions can be readily applied. Another example is the stochastic time-dependent current density functional theory method,[48-53] where the effects of the electronic reservoirs are modeled by an effective bath induced fluctuation term and a compensating dissipation term added to the Hamiltonian of the finite system.

An alternative approach, developed by Car, Burke, and Gebauer, does not attempt to model the semi-infinite leads, but rather replaces them with a single lead setup subject to ring boundary conditions.[54-58] Here, electrons that exit the simulation box on one side of the model junction re-enter at the other side. Due to the periodic nature of the method, the electrons, which are accelerated by an external field, have to be decelerated to avoid unrealistic velocity buildup. This is done via the introduction of an electron-phonon coupling term in the density matrix quantum master equation formulation of the problem. Unfortunately, computational demands dictate the use of relatively small lead models and thus the temperature of the phonon bath and the coupling strength have to be tuned to unphysically large values to obtain reasonable transport results.[59] Despite this, the intuitive nature of such density matrix based approaches, as well as their relative simplicity in terms of practical implementation, have triggered many studies in recent years aiming to harness their advantages toward advanced modeling of molecular electronics scenarios.[28, 60-67]

Going through the hierarchy of approximate methods, probably the conceptually simplest approach completely avoids the challenge of mimicking the semi-infinite leads by explicitly considering a closed finite system. A straightforward application of this idea is the microcanonical approach of Di Ventra and Todorov,[68] in which the discharge dynamics of two finite electron reservoirs contacted through a nanoscale/molecular bridge is investigated.[69] When the reservoirs are taken to be sufficiently large, a quasi-steady state can develop, thus allowing the study of both transient effects and the formation of the quasi-steady state current.[70] We note that a related method, based on constrained-DFT formalism, was suggested by Van Voorhis and co-workers.[71-73] A major advantage of these methods is that they are formally exact for closed systems, while enabling the simulation of relatively large molecular junction models. However, they are only suitable for relatively short simulation times when no backscattered electrons from the boundaries of the finite system enter the active region. Furthermore, Ercan et al.[74] have recently generalized the microcanonical approach to treat degenerate systems and examined its limitations using simple Tight-Binding (TB) models. It was concluded that the size, symmetry, and dimensionality of the system should be carefully taken into account when applying this methodology.

The absence of a true steady-state in the microcanonical picture was addressed by Sánchez et al.[75] They augmented the equations of motion with a driving term, acting near the edges of the finite electrode models so as to maintain a charge imbalance between the two finite reservoirs. This method successfully produces a long-lasting steady-state that resembles in nature the quasi-steady-state developing in microcanonical simulations. Furthermore, it is conceptually simple and requires merely a straightforward extension of standard electron-dynamics simulation techniques in closed-systems scenarios, thus enabling the treatment of relatively large junction models. Nevertheless, in the original formulation, some important limitations remain. These include: (i) a non-unique definition of the bias voltage imposed on the junction; (ii) a non-thermal distribution within the finite electronic reservoir models; (iii) complications in treating non-linear junction models including multi-lead configurations; and (iv) deviation of the density matrix from the $N$-representability condition.[76] We note that all but the latter limitation were, in principle, resolved in a related method that was developed for phenomenological models of molecular electronic junctions by Subotnik et al.[77] Nevertheless, this model addresses general phenomena of electronic conductance and does not take into account the detailed atomistic description of specific molecular junctions.

In the present study, we develop a new approach for time-domain simulations of electronic transport in molecular junctions with open boundary conditions. The new method is based on the driven Liouville-von Neumann equation suggested by Sánchez et al.[75] and Subotnik et al.,[77] with some important modifications to the working equation that provide an appropriate description of the system-bath damping. Using a unique transformation of the Hamiltonian from a real-space (site) representation to a state representation of the various sections of the system, we are able to mitigate the limitations discussed above while maintaining a fully atomistic description of the model junction. This bridges directly between atomistic and phenomenological models of molecular electronics junctions and thus allows harnessing the advantages of both representations of the problem.

The article is organized as follows: In section 2, we present a detailed account of our methodology. Section 3 is devoted to evaluating the performance of the suggested method for several model systems, including a linear tight-binding chain and an asymmetric three-terminal setup. Here, the ability to simulate bias-voltage and thermo-voltage effects is demonstrated by comparison of the obtained steady-state currents to the corresponding results calculated using the Landauer formalism. In section 4, we summarize and provide an outlook.

## 2. Methodology

a. Working Equation

Our starting point is the driven Liouville von-Neumann equation, developed by Sánchez *et al.* for electronic transport calculations.[75] As discussed above, within this method the molecular junction is represented by a finite system consisting of two sufficiently large lead models bridged by an active (extended-)molecule. The open boundary conditions are enforced by continuously driving the density matrix at the far edges of the lead models towards a charge polarized state, with charge accumulation on one side of the molecular junction and charge depletion on the other side. This is done via a driving term added to the Liouville-von Neumann equation in the following manner (unless mentioned otherwise atomic units are used throughout the paper):

$$\frac{d\hat{\rho}}{dt} = -i[\hat{H}, \hat{\rho}] - \Gamma(\hat{\rho} - \hat{\rho}^0), \quad (1)$$

where $\hat{H}$ is the matrix representation of the Hamiltonian of the system, $\hat{\rho}$ is the one-electron reduced density matrix, $\Gamma$ is a real number, which serves as a driving rate, and $\hat{\rho}^0$ is a target density matrix, defined in terms of its matrix elements in some real-space or atomic centered basis set representation. In the original formulation of the theory, the target density matrix was defined as follows:

$$\hat{\rho}_{ij}^0 = \begin{cases} \hat{\rho}_{ij}(t_0) & i,j \in C \\ \hat{\rho}_{ij}(t) & \text{otherwise.} \end{cases} \quad (2)$$

Here, C represents a group of indices of either real space grid points or atomic centered orbitals belonging to a predefined region in space in the vicinity of the far edges of the lead models, in which charge polarization is enforced. $\hat{\rho}(t_0)$ represents a density matrix encoding a physical state of charge polarization at the edges of the system. For a two-lead setup, C represents the left and right finite lead models and the equation, in matrix form, translates to:

$$\frac{d}{dt}\begin{pmatrix} \hat{\rho}_L & \hat{\rho}_{L,EM} & \hat{\rho}_{L,R} \\ \hat{\rho}_{EM,L} & \hat{\rho}_{EM} & \hat{\rho}_{EM,R} \\ \hat{\rho}_{R,L} & \hat{\rho}_{R,EM} & \hat{\rho}_R \end{pmatrix} =$$
$$-i\left[\begin{pmatrix} \hat{H}_L & \hat{V}_{L,EM} & \hat{0} \\ \hat{V}_{EM,L} & \hat{H}_{EM} & \hat{V}_{EM,R} \\ \hat{0} & \hat{V}_{R,EM} & \hat{H}_R \end{pmatrix}, \begin{pmatrix} \hat{\rho}_L & \hat{\rho}_{L,EM} & \hat{\rho}_{L,R} \\ \hat{\rho}_{EM,L} & \hat{\rho}_{EM} & \hat{\rho}_{EM,R} \\ \hat{\rho}_{R,L} & \hat{\rho}_{R,EM} & \hat{\rho}_R \end{pmatrix}\right] - \Gamma\begin{pmatrix} \hat{\rho}_L - \hat{\rho}_L^0 & \hat{0} & \hat{\rho}_{L,R} \\ \hat{0} & \hat{0} & \hat{0} \\ \hat{\rho}_{R,L} & \hat{0} & \hat{\rho}_R - \hat{\rho}_R^0 \end{pmatrix}, \quad (3)$$

where the Hamiltonian and density matrices are formally divided into the left (*L*), extended molecule (*EM*), and right (*R*) blocks (see Fig. 1), the direct coupling between the leads is assumed to be negligibly small, and $\hat{\rho}_L^0$, and $\hat{\rho}_R^0$ are the density matrices of the charged left and right lead models of the system, respectively. A similar equation was also used in Ref. [77] to study steady-state solutions of the time-dependent equation.

In this equation, the diagonal driving terms $\Gamma\hat{\rho}_{L/R}^0$ may be viewed as source terms injecting electrons into the extended molecule region, the diagonal $-\Gamma\hat{\rho}_{L/R}$ terms serve as absorbing sinks for electrons within the finite lead models and as damping terms for intra-lead coherences, and the corresponding off-diagonal terms $-\Gamma\hat{\rho}_{L,R/R,L}$ damp the inter-lead coherences. Here, both the EM block and the EM-lead coherences blocks of the density matrix remain undamped. We recall, however, that a microscopic derivation of the quantum master equation indicates that if the population in state i is damped with a rate $\gamma_i$ and that of state j is damped with a rate $\gamma_j$ then the contribution of this population relaxation to the relaxation of their mutual coherence is given by the so called $t_1$ expression: $\frac{1}{2}(\gamma_i + \gamma_j)$.[78, 79] Hence, if the eigenstates of the lead models are damped with a rate $\Gamma$ then the lead-lead coherences should be damped with the same rate, as done in Eq.

(3), but the lead-EM coherences should be damped, as well, with a rate of $\frac{1}{2}\Gamma$. Therefore, the resulting driven Liouville von-Neumann equation should read:

$$\frac{d}{dt}\begin{pmatrix} \hat{\rho}_L & \hat{\rho}_{L,EM} & \hat{\rho}_{L,R} \\ \hat{\rho}_{EM,L} & \hat{\rho}_{EM} & \hat{\rho}_{EM,R} \\ \hat{\rho}_{R,L} & \hat{\rho}_{R,EM} & \hat{\rho}_R \end{pmatrix} =$$

$$-i\left[\begin{pmatrix} \hat{H}_L & \hat{V}_{L,EM} & \hat{0} \\ \hat{V}_{EM,L} & \hat{H}_{EM} & \hat{V}_{EM,R} \\ \hat{0} & \hat{V}_{R,EM} & \hat{H}_R \end{pmatrix}, \begin{pmatrix} \hat{\rho}_L & \hat{\rho}_{L,EM} & \hat{\rho}_{L,R} \\ \hat{\rho}_{EM,L} & \hat{\rho}_{EM} & \hat{\rho}_{EM,R} \\ \hat{\rho}_{R,L} & \hat{\rho}_{R,EM} & \hat{\rho}_R \end{pmatrix}\right] - \Gamma\begin{pmatrix} \hat{\rho}_L - \hat{\rho}_L^0 & \frac{1}{2}\hat{\rho}_{L,EM} & \hat{\rho}_{L,R} \\ \frac{1}{2}\hat{\rho}_{EM,L} & \hat{0} & \frac{1}{2}\hat{\rho}_{EM,R} \\ \hat{\rho}_{R,L} & \frac{1}{2}\hat{\rho}_{R,EM} & \hat{\rho}_R - \hat{\rho}_R^0 \end{pmatrix}, \qquad (4)$$

Eq. (4) serves as our working equation throughout this article. A detailed heuristic derivation of it can be found in appendix A. Importantly, although both Eq. (3) and Eq. (4) do not forbid it,[76] in practice we find neither violations of Pauli's exclusion principle nor deviations from density matrix positivity throughout all tight-binding based simulations presented below, using Eq. (4). However, both problems are observed in the same calculations if Eq. (3) is used instead. Furthermore, use of Eq. (4) results in much less noise in the time-dependent current and a considerably smoother and faster convergence of the current towards steady-state, than the equivalent results obtained using Eq. (3).

b. Target Density

In the original formulation of Sánchez *et al.* the target density, $\hat{\rho}(t_0)$, was obtained from a ground state calculation of the same system under the presence of an external electric field, parallel to the axial direction of the junction.[75] The application of the field causes charge separation and accumulation at the two edges of the system, thus aiming to mimic the case of charge imbalance due to the application of an external bias under non-equilibrium conditions. Once $\hat{\rho}(t_0)$ is obtained the field is switched off and the density matrix, which no longer represents a stationary state, is allowed to evolve according to Eq. (3), i.e., subject to the (soft) constraint that its value at the edge regions $(i, j \in C)$ should approach the value of $\hat{\rho}(t_0)$. In this way, in the extended molecule region, far from the edges of the lead models $(i, j \notin C)$, the damping term on the right hand side of Eq. (3) vanishes and the density matrix evolves according to the standard Liouville-Von Neumann equation. Near the boundaries $(i, j \in C)$ of the lead models, the damping term constantly drives the density matrix towards the reference value, $\hat{\rho}_{ij}(t_0)$ thus enforcing charge polarization close to the edges. The driving rate $\Gamma$ could, in principle, be derived from the properties of the metallic leads. However, practically, it served as a free parameter arbitrarily chosen within reasonable physical bounds to produce a stable steady state. If the damping factor is chosen to be identically zero throughout space ($\Gamma = 0$) then the microcanonical picture is restored.

In the above approach, the initial conditions and target densities are enforced via an external electric field that induces a charge imbalance, which resembles a scenario, where a capacitor is discharged through the molecular junction. This, however, does not provide a unique definition of the chemical potentials of the leads and hence the bias voltage, as well as the thermal distribution of the electronic occupations. Furthermore, since the charge separation is induced by an external uniform electric field, the finite lead models have to be aligned along the field axis, resulting in a geometry that inhibits the addition of further electrodes to the model system. The latter problem can, in principle, be solved by building target densities based on more complex electric fields or constrained DFT.[71] However, these considerably complicate the procedure and do not provide a solution to the other important issues raised above.

Importantly, when formulating the problem in the basis of the states of the separate leads and extended molecule subsystems, all the above mentioned problems are absent and a clear and unique definition of both the bias voltage and the electronic temperature exists.[77] Thus, if one desires to maintain the detailed atomistic information of the junction while avoiding the problems associated with the appropriate definition of the boundary conditions, it is required to find a transformation from the atomistic representation of the system used by Sánchez et al.[75] to the state representation considered by Subotnik et al.[77]

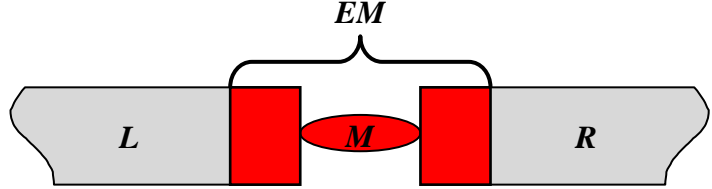

Figure 1: A schematic representation of the molecular junctions divided into its three parts: (i) Left lead (L); (ii) Right lead (R); and (iii) Extended molecule (EM) marked in red, which consists of the molecule (M) and its adjacent lead sections.

To this end, we use the same formal division presented in Eqs. (3) and (4) of the molecular junction into three sections: (i) the left lead (L); (ii) the right lead (R); and (iii) the extended molecule (EM; see Fig. 1). As mentioned above, the latter is the molecule, augmented by some portion of the leads such that near the far edges the effect of the interface with the molecule on the electronic structure of the lead sections becomes marginal. With this somewhat arbitrary division, the localized basis set or real-space matrix representation of the Hamiltonian operator ($\hat{H}$) obtains the following form (referred to below as the "atomistic-" or "site-representation"):

$$\hat{H} = \begin{pmatrix} \hat{H}_L & \hat{V}_{L,EM} & \hat{0} \\ \hat{V}_{EM,L} & \hat{H}_{EM} & \hat{V}_{EM,R} \\ \hat{0} & \hat{V}_{R,EM} & \hat{H}_R \end{pmatrix}. \quad (5)$$

Here, $\hat{H}_i$ is the Hamiltonian matrix block of the $i$th section of the system and $\hat{V}_{ij}$ represents the interactions between section $i$ and section $j$ of the system where $i,j=(L, EM, R)$. If the interactions are short ranged $\hat{V}_{ij}$ becomes a sparse matrix and, as mentioned above, we can safely assume that there exist no direct coupling between the two leads $\hat{V}_{L,R} = \hat{V}_{R,L} = 0$.

We now define the transformation of this Hamiltonian matrix representation to the basis of eigenfunctions of the individual isolated system sections (*L*, *EM*, and *R*). Denoting by $\hat{U}_i$ the unitary transformation matrix that diagonalizes the $\hat{H}_i$ block such that $\hat{U}_i^\dagger \hat{H}_i \hat{U}_i = \widetilde{\hat{H}}_i$, where $\widetilde{\hat{H}}_i$ is a diagonal square matrix with the eigenenergies of the isolated $i$th section on its diagonal, we can write the following global unitary transformation matrix:

$$\hat{U} = \begin{pmatrix} \hat{U}_L & \hat{0} & \hat{0} \\ \hat{0} & \hat{U}_{EM} & \hat{0} \\ \hat{0} & \hat{0} & \hat{U}_R \end{pmatrix}, \tag{6}$$

which transforms the full Hamiltonian matrix of Eq. (3) in the following manner:

$$\widehat{\widetilde{H}} = \hat{U}^\dagger \hat{H} \hat{U} = \begin{pmatrix} \widehat{\widetilde{H}}_L & \widehat{\widetilde{V}}_{L,EM} & \hat{0} \\ \widehat{\widetilde{V}}_{EM,L} & \widehat{\widetilde{H}}_{EM} & \widehat{\widetilde{V}}_{EM,R} \\ \hat{0} & \widehat{\widetilde{V}}_{R,EM} & \widehat{\widetilde{H}}_R \end{pmatrix}. \tag{7}$$

Here, we have defined the matrix that couples between the eigenstates of the *i*th section and those of the *j*th section as $\widehat{\widetilde{V}}_{ij} \equiv \hat{U}_i^\dagger \hat{V}_{ij} \hat{U}_j$. This representation, which shall be referred to as the "state representation", represents three separate sets of quantum states (of the left lead, extended molecule, and the right lead) that are coupled according to the coupling scheme presented in Fig. ure 2. Note that the fact that we have neglected the direct coupling between the leads in the site representation is reflected in the state representation as well.

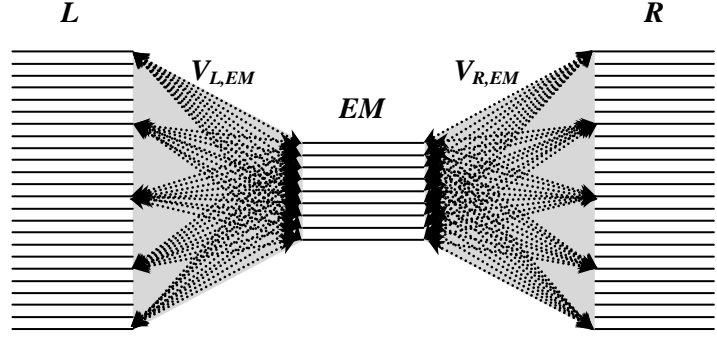

Figure 2: A scheme of the transformation to the state representation where the manifold of eigenstates of the extended molecule couples separately to the manifolds of eigenstates of the left and right leads.

Similarly, the reduced density matrix transforms as follows:

$$\hat{\widetilde{\rho}} = \hat{U}^\dagger \hat{\rho} \hat{U} = \begin{pmatrix} \hat{\widetilde{\rho}}_L & \hat{\widetilde{\rho}}_{L,EM} & \hat{\widetilde{\rho}}_{L,R} \\ \hat{\widetilde{\rho}}_{EM,L} & \hat{\widetilde{\rho}}_{EM} & \hat{\widetilde{\rho}}_{EM,R} \\ \hat{\widetilde{\rho}}_{R,L} & \hat{\widetilde{\rho}}_{R,EM} & \hat{\widetilde{\rho}}_R \end{pmatrix}, \tag{8}$$

with $\hat{\widetilde{\rho}}_{i,j} = \hat{U}_i^\dagger \hat{\rho}_{i,j} \hat{U}_j$. Hence, in the state-representation Eq. (4) assumes the following form (see Appendix A):

$$\frac{d}{dt}\hat{\widetilde{\rho}} = -i\left[\widehat{\widetilde{H}}, \hat{\widetilde{\rho}}\right] - \Gamma \begin{pmatrix} \hat{\widetilde{\rho}}_L - \hat{\widetilde{\rho}}_L^0 & \frac{1}{2}\hat{\widetilde{\rho}}_{L,EM} & \hat{\widetilde{\rho}}_{L,R} \\ \frac{1}{2}\hat{\widetilde{\rho}}_{EM,L} & \hat{0} & \frac{1}{2}\hat{\widetilde{\rho}}_{EM,R} \\ \hat{\widetilde{\rho}}_{R,L} & \frac{1}{2}\hat{\widetilde{\rho}}_{R,EM} & \hat{\widetilde{\rho}}_R - \hat{\widetilde{\rho}}_R^0 \end{pmatrix}, \tag{9}$$

where the diagonal terms represent the eigenstate occupations of the three isolated sections of the system. Thus, we may now choose as our target density matrix blocks, $\hat{\widetilde{\rho}}_{L/R}^0$, diagonal sub-matrices that represent the equilibrium state of the isolated left and right leads in energy space and are populated according to the equilibrium Fermi-Dirac distribution of the respective lead, $f_{L/R}(E_n^{L/R}) = 1/\left[e^{(E_n^{L/R} - \mu_{L/R})/(k_B T_{L/R})} + 1\right]$, where $E_n^{L/R}$ are the eigenenergies of the finite left and right lead model, $T_{L/R}$ are the electronic temperatures of the left and right leads, $k_B$ is Boltzmann's constant, and $\mu_{L/R}$ are the chemical potential of the left and right leads set by the bias voltage, V, such that $\mu_{L/R} = E_F^{L/R} \pm 0.5V$, and $E_F^{L/R}$ are the Fermi energies of the left and right leads, respectively. Assuming that, within the energy bandwidth of the finite lead models, the lead densities of states are sufficiently dense, this new definition of the target density provides a physically sound representation of the boundary conditions, which has no ambiguity with respect to the definition of the bias voltage (given by the difference in chemical potentials of the leads) and can take into account electronic thermal effects. Furthermore, this procedure is not limited to two lead setups and can be readily applied to the case of multiple lead junctions.

Within this scheme the actual calculation is performed according to the following steps: (i) **Geometry definition**: Definition of the geometry of the molecular junction including two (or more) finite, and sufficiently large, lead models and a bridging (extended-)molecule; (ii) **Transformation to the state representation**: Diagonalization of the separate system sections, construction of the global transformation matrix, and transformation of the full Hamiltonian matrix from the atomistic to the state representation; (iii) **Target density formation**: Form the target density via the population of the various leads states according to their respective chemical potential (bias voltage) and temperature; (iv) **Initial state**: Construction of the initial state where the lead sections are populated at their target density values and the extended molecule is populated up to the Fermi energy. This represents a separated system in which the couplings are turned on at time zero. Naturally, other initial states can be considered; (v) **Propagation**: Propagation of the density matrix according to Eq. (4) with a given damping factor. The time dependent current density is then monitored during the simulation.

## 3. Results and Discussion

We start by applying the suggested methodology to a model two-lead molecular junction, where the leads are represented by two semi-infinite atomic chains and the extended molecule is represented by a finite atomic chain coupled, locally, to both lead models (see Fig. 3). For the time propagation of the density matrix we use the fourth order Runge-Kutta scheme throughout this study.[80] To represent the electronic structure of the system we choose a tight-binding model. We believe that this model is simple enough to provide a clear picture of the performance of the suggested approach, while avoiding complications arising from more involved molecular junction models and more complex electronic structure methods. It should be emphasized, however, that our time-dependent transport methodology is not limited to tight-binding representations and it can be applied with more advanced electronic structure methods.

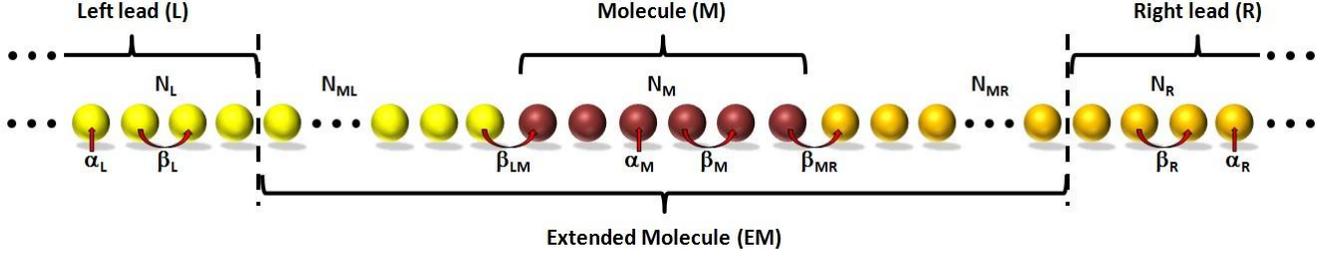

Figure 3: Schematic representation of the tight-binding two-lead model. Yellow, maroon, and orange spheres represent the left lead, molecule, and right lead, respectively. The extended molecule region is marked explicitly. $\alpha_L$, $\alpha_M$, $\alpha_R$, $\beta_L$, $\beta_M$, and $\beta_R$ mark the onsite energies ($\alpha$) and hopping integrals ($\beta$) of the left lead (L), molecule (M), and right lead (R) subsystems, respectively. $\beta_{LM}$ and $\beta_{MR}$ are the coupling matrix elements between the left lead and the molecule and between the molecule and the right lead, respectively. $N_L$, $N_M$, and $N_R$ are the number of sites used to represent the left lead, molecule, and right lead models, respectively. $N_{ML}$ and $N_{MR}$ are the number of extended molecule atoms belonging to the left and right leads, respectively.

First, we study the effects of the choice of driving rate ($\Gamma$) on the time-dependent current, *I(t)*, calculated in the atomistic/site-representation as the bond current at the center of the molecular bridge model using the relation $I_{n,n+1}(t) = (2\beta e/\hbar) Im[\rho_{n,n+1}]$ (see Appendix B). Here, *e* is the electron charge, $\hbar$ is the reduced Planck constant, $\beta$ is the hopping matrix element between site *n* and site *n+1*, and $\rho_{n,n+1}$ is the off-diagonal element of the density matrix representing the relevant site coherences.[81] As can be seen in Fig. 4 for a uniform tight-binding chain, in the microcanonical case ($\Gamma$=0) we obtain the expected behavior, where the initial transient current oscillations gradually develop into a quasi-steady-state (QSS) that matches well the steady-state current value predicted by the Landauer approach and persists until the propagating electronic wavepacket reaches the boundaries of the finite model system (~1,060 fs in our model). At this time, the wavepacket is reflected back towards the bridge causing the current to reverse its sign. Because no damping is introduced, this process is repeated in a quasi-periodic manner.

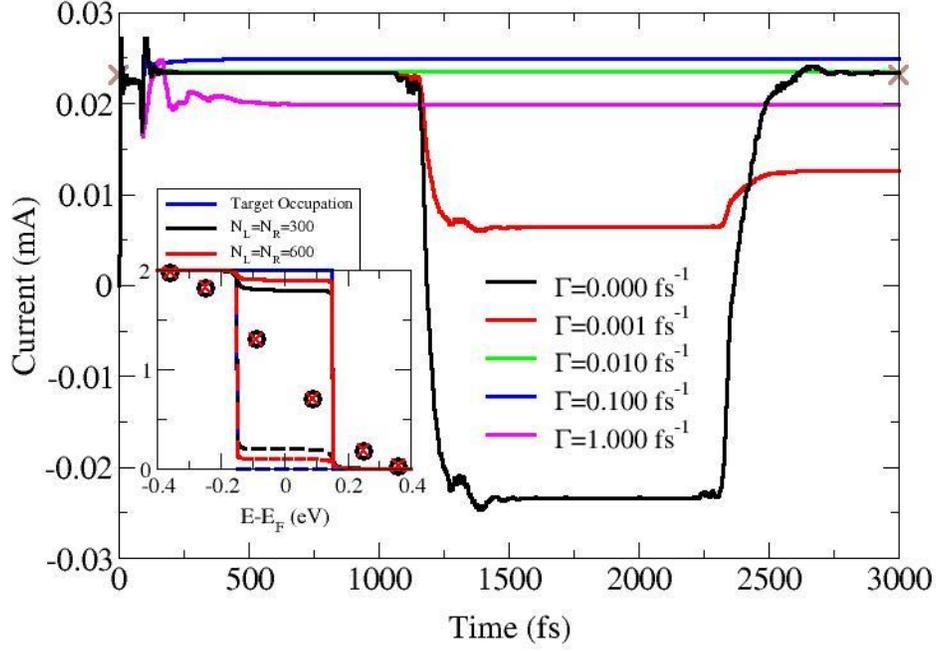

Figure 4: Effect of the driving rate ($\Gamma$) on the time-dependent current calculated for a tight-binding atomic chain model (see Fig. 3) under a bias voltage of $V_b$=0.3 V and electronic lead temperatures of $T_L=T_R$=0K. In these calculations, the model dimensions are chosen to be $N_L=N_R$=300, $N_{ML}=N_{MR}$=50, and $N_M$=6, the on-site energies are taken as $\alpha_L=\alpha_M=\alpha_R$=0 eV, and the hopping integrals used are $\beta_L=\beta_M=\beta_R=\beta_{LM}=\beta_{MR}$=-0.2 eV. The black curve was obtained using the microcanonical approach ($\Gamma$=0) in the state representation. The red, green, blue, and purple curves were obtained using the methodology suggested here with $\Gamma$=0.001, 0.01, 0.1, and 1.0 fs$^{-1}$, respectively. The brown X marks represent the steady-state current obtained via the Landauer approach (see Appendix C). A time step of 1 fs was used throughout the simulations. Inset: Left lead (full lines), right lead (dashed lines), and molecule (symbols) steady-state occupations obtained using lead models of $N_L=N_R$=300 (black) and $N_L=N_R$=600 (red) compared to the corresponding target lead-equilibrium step-function distributions (blue).

When a finite, but too small, driving rate is chosen ($\Gamma$=0.001 fs$^{-1}$, red curve in Fig. 4), back-scattering due to edge reflections is somewhat suppressed. Similar to the microcanonical case, an initial QSS develops with a current that matches the Landauer value. When the wavepacket reaches the boundaries, this QSS is destroyed and the current reduces due to the remaining backscattering. Nevertheless, unlike the case of the microcanonical simulation, a complete current reversal is not observed and, following some further back and forward scattering from the finite model edges, a stable steady-state develops. Importantly, the obtained steady-state current is considerably smaller than the initial QSS and the Landauer steady-state values, indicating that indeed the driving rate is too small. On the other hand, when the damping term is taken to be too large the initial dynamics deviate from the microcanonical behavior and the obtained steady-state is either larger ($\Gamma$=0.1 fs$^{-1}$, blue curve) or smaller ($\Gamma$=1.0 fs$^{-1}$, purple curve) than the Landauer value.

The behavior described above for extreme driving rate values, which has been previously rationalized in Ref. [75], suggests that by fine tuning of the damping rate a stable steady state should be obtainable. Indeed, when choosing a rate of $\Gamma=0.01$ fs$^{-1}$ (green curve in Fig. 4) a stable steady-state occurs, which matches well the Landauer value.[82] The value of this steady-state current is slightly larger than the initial QSS current of the microcanonical simulation. This can be rationalized by the fact that when the microcanonical QSS is achieved, the chemical potential difference between the finite lead models reduces with respect to the initial state due to the discharge dynamics. In the case of the driven equation, the leads are kept close to their desired equilibrium distribution throughout the simulation, including at steady-state.

An important observation made in Fig. 4 is that the obtained steady-state current is quite stable with respect to the choice of driving rate. Here, changing $\Gamma$ by three orders of magnitude results in a steady-state that varies between 0.011-0.025 mA with our model parameters. Some physical insights regarding the value of the optimal driving rate were given in the original paper of Sánchez et al in terms of the tight-binding hopping integral and surface local density of states.[75] We note that the damping of the backscattering should occur at a rate that is of the order of the timescale that it takes the wavepacket to be reflected from the boundaries of the simulation box. Evaluating this time-scale as the current switching time obtained in the microcanonical simulation ($\Delta t \sim 100$ fs) we obtain an estimated driving rate of $\Gamma \sim 1/\Delta t \sim 0.01$ fs$^{-1}$, which is indeed the value used above to produce a steady-state current that matches well both the Landauer value and the QSS of the microcanonical simulation.

In order to better understand the nature of the steady state obtained from our finite model system and its relation to the Landauer picture, we plot in the inset of Fig. 4 the steady-state eigenstate occupations of the molecule and the left and right lead models for two lead sizes and compare them to the corresponding Fermi-Dirac lead equilibrium distributions used in the Landauer formalism. As can be seen, the steady-state occupations in the left and right leads (black lines) somewhat deviate from the target occupation function (blue curves) within the Fermi bias transport window of $\pm 0.15$ eV. This results from the fact that with the given finite number of electrons it is impossible to simultaneously fix the electronic distributions in the finite lead models at their equilibrium form with appropriate chemical potential and electronic temperature and obtain steady-state dynamics in the molecular section.[77] This, however, can be improved by increasing the size of the lead models and thus also the number of electrons in the system. Indeed, the red curves in the inset of Fig. 4, which were obtained by doubling the lead model sizes from 300 to 600 sites, show much smaller deviations from the target densities at the lead models. Importantly, the effect of these deviations on the steady-state occupation of the molecular states (black circles and red x-marks) that, owing to the bias voltage, is different from their equilibrium distribution,[78, 83-85], as well as on the calculated current (~0.16%), is marginal.

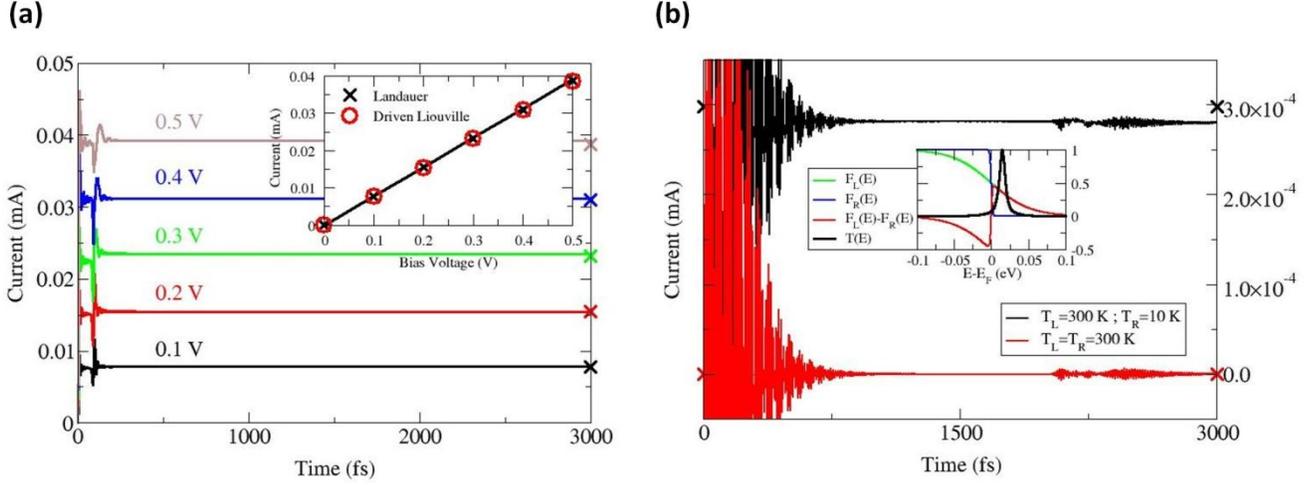

Figure 5: Bias- and thermo-voltage simulations carried out for the system depicted in Fig. 3. (a): Time-dependent current calculated at various bias voltages, using the same system parameters given in the caption of Fig. 4 with a driving rate of $\Gamma=0.01$ fs$^{-1}$. Colored x marks designate the corresponding steady-state currents calculated via the Landauer formalism. Inset: current vs. bias curve calculated from the stead-state currents obtained at a simulation time of 3 picoseconds (red circles) and the Landauer formalism (black X marks). (b): Time-dependent currents calculated for lead temperature differences of $\Delta T=290$ K (black curve) and 0 K (red curve) at zero bias voltage. The model dimensions are $N_L=N_R=600$, $N_{ML}=N_{MR}=50$, and $N_M=6$, the on-site energies are taken as $\alpha_L=\alpha_R=0$ eV and $\alpha_M=-0.075$ eV, and the hopping integrals used are $\beta_L=\beta_M=\beta_R=-0.2$ eV and $\beta_{LM}=\beta_{MR}=-0.04$ eV. The driving rate is fine-tuned to a value of $\Gamma=0.0025$ fs$^{-1}$, which yields good agreement with the Landauer steady-state currents (x marks) and a stable steady-state. Inset: Transmittance probability (black), left (green) and right (blue) lead Fermi distribution functions, and the Fermi transport window (red) of the Landauer formalism. A time step of 1 fs is used throughout the simulations.

We now turn to discuss how bias- and thermo-voltage effects can be readily investigated using the proposed method. As mentioned above, via the target edge density matrices, which encode the required electronic occupation functions of the leads, our new approach provides an unambiguous definition of the electronic temperature and bias voltage applied on the junction and assures that they remain constant (if desired) throughout the simulation. To demonstrate this, we present, in the left panel of Fig. 5, the time-dependent current through a 706 atom chain (see Fig. 3) for various bias voltages. As may be expected for the system investigated, the steady-state current increases with increasing bias voltage. Furthermore, the current-voltage characteristics obtained by extracting the steady-state currents from the dynamical simulation and those calculated via the Landauer formalism are in excellent agreement, both predicting an Ohmic behavior within the bandwidth of the lead models (see inset of the left panel).

In the right panel of Fig. 5, we consider the case of thermo-voltage, where electrical currents are induced via an electronic temperature gradient held constant between the leads.[86, 87] To this end, we consider the case of low coupling, where the lead-molecule hopping matrix elements are taken to be 20% of the hopping integrals within the leads and molecule chain models. The system is gated in order to shift one of the transmission resonances to an asymmetric position around the Fermi energy, such that the positive (electronic conduction) part of the Fermi transport window encompasses it (see inset of Fig. 5b), resulting in a finite electronic current that flows through the system from the warm to the cold lead. In the dynamical simulations, depicted in the main panel, after an initial period of relatively strong oscillations, the time dependent current (black curve) relaxes towards a steady-state that is in good agreement with the Landauer value (black curve). We note that the relative difference between the steady-state currents obtained using the two methods in the present case (~6%) is somewhat larger than that shown so far. However, since the overall currents obtained here are quite small, in terms of the absolute values (~0.017 µA) this is a minor deviation. For comparison, we present the time-dependent current in the absence of temperature gradient and bias voltage (red curve) where, as may be expected, the steady-state current vanishes. The short-time oscillations obtained in this case are a result of the initial conditions where the three subsystems are disconnected and set to their own equilibrium distribution. The increased noise appearing in both curves at ~2 ps results from residual backscattering occurring due to the relatively small driving rate used in these simulations. Similar recurrences occur at longer simulation times. However, their amplitude decreases rapidly to provide a stable steady-state.

Finally, we demonstrate how our method can be readily applied to more complex molecular junctions such as multi-lead configurations. Such junctions are very appealing in the realm of molecular electronics as they enable the design of molecular electronics components presenting novel functionalities that take advantage of coherent transport effects.[64, 88-91] As shown below, a suitable choice of the target density allows for modeling several source/sink reservoirs that couple to the molecular bridge. To demonstrate this capability, we choose the three-lead junction depicted in the right panel of Fig. 6, where the upper lead (UL) serves as a source and the left (LL) and right (RL) leads serve as sinks. As in the two-lead setup discussed above, the three semi-infinite leads are represented by finite atomic chains (marked by green, yellow, and orange spheres in the right panel) that couple locally to the three arms of the molecule model. A buffer region of the first 50 lead sites, adjacent to each arm of the molecule model, constructs the extended-molecule beyond which the driving terms act. The left and right arms of the molecule model do not couple directly and the coupling of the upper arm to the right arm is taken to be half the coupling to the left arm, thus forming an asymmetrically coupled T-junction.

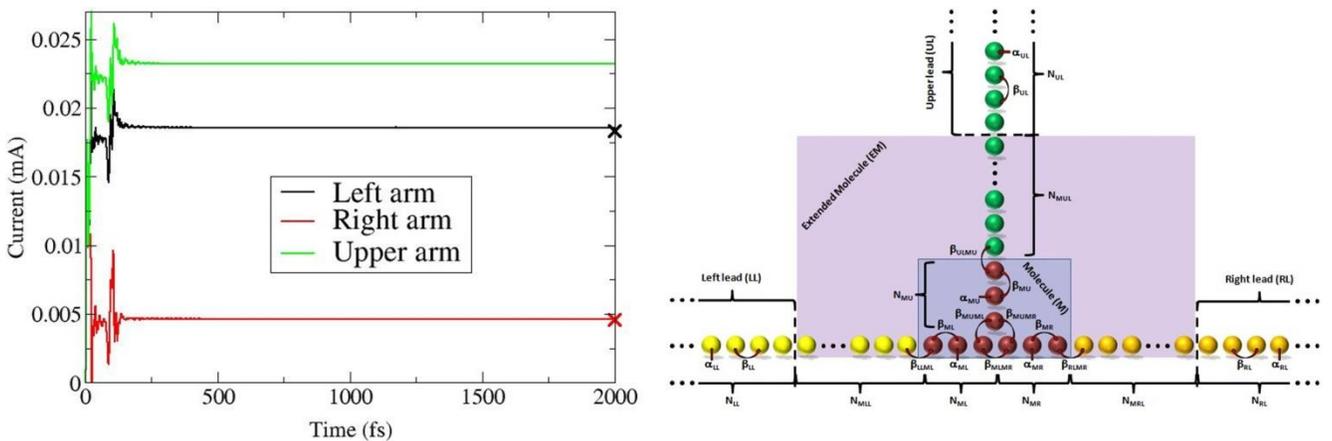

Figure 6: Time-dependent current for a three-lead tight-binding T-junction under a bias voltage of $V_b = 0.3$ V with a higher chemical potential of $\mu_U = E_F + \frac{1}{2}eV_b$ at the upper lead and equal lower chemical potentials of $\mu_L = \mu_R = E_F - \frac{1}{2}eV_b$ at the left and right leads, and an electronic temperature of 0K for all leads. A driving rate of $\Gamma$=0.01 fs$^{-1}$ and a time step of 1 fs are used. The junction parameters (see right panel) are as follows: $N_{ML} = N_{MR} = N_{MU} = 6$, $N_{LL} = N_{RL} = N_{UL} = 300$, $N_{MLL} = N_{MRL} = N_{MUL} = 50$, $\alpha_{ML} = \alpha_{MR} = \alpha_{MU} = \alpha_{LL} = \alpha_{RL} = \alpha_{UL} = 0\ eV$, $\beta_{ML} = \beta_{MR} = \beta_{MU} = \beta_{LL} = \beta_{RL} = \beta_{UL} = \beta_{LLML} = \beta_{RLMR} = \beta_{ULMU} = \beta_{MUML} = -0.2\ eV$, $\beta_{MUMR} = -0.1\ eV$, $\beta_{MLMR} = 0\ eV$. The bond currents are calculated at the center of the left (black curve), right (red curve), and upper (green curve) molecular arms. Black and red X marks represent the Landauer currents in the left and right arms, respectively.

The formalism described above can be readily extended to treat such a multi-lead setup. Here, the Hamiltonian matrix in the atomistic/site representation is given by:

$$\hat{H} = \begin{pmatrix} \hat{H}_{EM} & \hat{V}_{EM,UL} & \hat{V}_{EM,LL} & \hat{V}_{EM,RL} \\ \hat{V}_{UL,EM} & \hat{H}_{UL} & \hat{0} & \hat{0} \\ \hat{V}_{LL,EM} & \hat{0} & \hat{H}_{LL} & \hat{0} \\ \hat{V}_{RL,EM} & \hat{0} & \hat{0} & \hat{H}_{RL} \end{pmatrix}, \quad (10)$$

where $\hat{H}_i$ is the Hamiltonian matrix block of the $i$th section of the system and $\hat{V}_{i,j}$ represents the coupling between section $i$ and section $j$ of the system where $i,j$=(EM, UL, LL, RL) and as before we assume no inter-lead couplings. The corresponding global transformation matrix to the state-representation is:

$$\hat{U} = \begin{pmatrix} \hat{U}_M & \hat{0} & \hat{0} & \hat{0} \\ \hat{0} & \hat{U}_{UL} & \hat{0} & \hat{0} \\ \hat{0} & \hat{0} & \hat{U}_{LL} & \hat{0} \\ \hat{0} & \hat{0} & \hat{0} & U_{RL} \end{pmatrix}, \quad (11)$$

where $\widehat{\tilde{H}}_i = \hat{U}_i^\dagger \hat{H}_i \hat{U}_i$ is a diagonal matrix holding on its diagonal the eigenstates of the isolated $i$th section of the system. The Hamiltonian of the full system in the state-representation is given by:

$$\widehat{\tilde{H}} = \hat{U}^\dagger \hat{H} \hat{U} = \begin{pmatrix} \widehat{\tilde{H}}_{EM} & \widehat{\tilde{V}}_{EM,UL} & \widehat{\tilde{V}}_{EM,LL} & \widehat{\tilde{V}}_{EM,RL} \\ \widehat{\tilde{V}}_{UL,EM} & \widehat{\tilde{H}}_{UL} & \hat{0} & \hat{0} \\ \widehat{\tilde{V}}_{LL,EM} & \hat{0} & \widehat{\tilde{H}}_{LL} & \hat{0} \\ \widehat{\tilde{V}}_{RL,EM} & \hat{0} & \hat{0} & \widehat{\tilde{H}}_{RL} \end{pmatrix}, \quad (12)$$

with $\widehat{\tilde{V}}_{i,j} = \hat{U}_i^\dagger \hat{V}_{i,j} \hat{U}_j$, and the driven Liouville von-Neumann equation in this representation is:

$$\frac{d}{dt}\widehat{\tilde{\rho}} = -i\left[\widehat{\tilde{H}}, \widehat{\tilde{\rho}}\right] - \Gamma \begin{pmatrix} \hat{0} & \frac{1}{2}\widehat{\tilde{\rho}}_{EM,UL} & \frac{1}{2}\widehat{\tilde{\rho}}_{EM,LL} & \frac{1}{2}\widehat{\tilde{\rho}}_{EM,RL} \\ \frac{1}{2}\widehat{\tilde{\rho}}_{UL,EM} & \widehat{\tilde{\rho}}_{UL} - \widehat{\tilde{\rho}}_{UL}^0 & \widehat{\tilde{\rho}}_{UL,LL} & \widehat{\tilde{\rho}}_{UL,RL} \\ \frac{1}{2}\widehat{\tilde{\rho}}_{LL,EM} & \widehat{\tilde{\rho}}_{LL,UL} & \widehat{\tilde{\rho}}_{LL} - \widehat{\tilde{\rho}}_{LL}^0 & \widehat{\tilde{\rho}}_{LL,RL} \\ \frac{1}{2}\widehat{\tilde{\rho}}_{RL,EM} & \widehat{\tilde{\rho}}_{RL,UL} & \widehat{\tilde{\rho}}_{RL,LL} & \widehat{\tilde{\rho}}_{RL} - \widehat{\tilde{\rho}}_{RL}^0 \end{pmatrix}. \quad (13)$$

Here, the state representation of the density matrix, $\widehat{\tilde{\rho}}$, is given by:

$$\hat{\tilde{\rho}} = \hat{U}^\dagger \hat{\rho} \hat{U} = \begin{pmatrix} \hat{\tilde{\rho}}_{EM} & \hat{\tilde{\rho}}_{EM,UL} & \hat{\tilde{\rho}}_{EM,LL} & \hat{\tilde{\rho}}_{EM,RL} \\ \hat{\tilde{\rho}}_{UL,EM} & \hat{\tilde{\rho}}_{UL} & \hat{\tilde{\rho}}_{UL,LL} & \hat{\tilde{\rho}}_{UL,RL} \\ \hat{\tilde{\rho}}_{LL,EM} & \hat{\tilde{\rho}}_{LL,UL} & \hat{\tilde{\rho}}_{LL} & \hat{\tilde{\rho}}_{LL,RL} \\ \hat{\tilde{\rho}}_{RL,EM} & \hat{\tilde{\rho}}_{RL,UL} & \hat{\tilde{\rho}}_{RL,LL} & \hat{\tilde{\rho}}_{RL} \end{pmatrix}, \tag{14}$$

and its various blocks can be expressed in terms of the corresponding blocks in the site representation $\hat{\rho}$ as $\hat{\tilde{\rho}}_{i,j} = \hat{U}_i^\dagger \hat{\rho}_{i,j} \hat{U}_j$. As in the case of the two-lead setup, the target densities $\hat{\tilde{\rho}}_{UL/LL/RL}^0$ are diagonal matrices representing the Fermi-Dirac equilibrium electron occupation distributions of the respective lead states encoding the appropriate chemical potential (taking into account the bias voltage) and electronic temperature.

The time-dependent bond currents in the three junction arms are depicted in the left panel of Fig. 6. Here, the incoming current from the upper lead is distributed between the left and right arms according to the respective couplings. Since the coupling of the upper arm to the left arm is chosen to be stronger than its coupling to the right arm, the bond current in the latter is consistently smaller than in the former. Importantly, the sum of the left and right (sink) arms steady-state bond currents equals the corresponding upper (source) arm counterpart, thus fulfilling Kirchhoff's first law. The good agreement between the steady-state bond currents, calculated at the sink leads, and the corresponding Landauer currents (x marks), calculated from the left and right transmittance probabilities, further supports the validity of our method.

## 4. Summary and Conclusions

We have presented a method for simulating electron dynamics in open quantum systems out of equilibrium, based on a driven Liouville von-Neumann equation approach with appropriate boundary conditions, applied to finite atomistic models. The Liouvillian operator describing the dynamics of the closed system is augmented with damping (sink) terms that serve to absorb electrons entering the finite lead models and to dephase the inter-lead, intra-lead, and lead-system coherences and driving (source) terms that inject electrons into the active region with the appropriate lead equilibrium electronic distribution. This is achieved by performing a transformation from the atomistic view of the system to a state representation, where the coupling scheme between the eigenstates of the various isolated sections of the full system is obtained explicitly. Unlike the atomistic representation, the state representation allows for an unambiguous definition of the bias voltage and lead electronic temperature. To demonstrate this, we have considered three tight-binding models: a homogeneous one-dimensional chain under an external bias voltage, a thermally biased weekly coupled chain, and a three-terminal junction. For all cases studied, while not strictly guaranteed by our formalism, positivity of the reduced density matrix was conserved throughout the simulations with no apparent deviations from Pauli's exclusion principle. Furthermore, steady-state Landauer results presented excellent agreement with the results of the dynamical model thus indicating the validity of the method. This opens the way for the study of many dynamical phenomena in molecular junctions, including the effects of alternating bias voltages and dynamical thermal effects, time-dependent electromagnetic fields, temporarily separated light-pulse induced electron dynamics, charge separation dynamics for photo-voltaic applications, transient currents, spin dynamics, and, with appropriate extensions, also coupled electron-nuclei dynamics. We therefore believe that the suggested method will contribute to the design of unique control schemes for charge and spin transport in single molecule junctions and will enable the discovery of novel structures that may lead to the development of new electronic components with diverse functionalities based on the promising concepts of molecular electronics and spintronics.

## Acknowledgements

We would like to thank Prof. Abraham Nitzan, Prof. Tamar Seideman, Prof. Mark Ratner, Prof. Roi Baer, Prof. Massimiliano Di Ventra, Prof. Uri Peskin, Dr. Yonatan Dubi, Prof. Daniel Neuhauser, and Prof. Michael Galperin for numerous helpful discussions. Work at TAU was supported by the German-Israeli Foundation under research grant No. 2291-2259.5/2011, the Israel Science Foundation under grant number 1740/13, the Lise-Meitner Minerva Center for Computational Quantum Chemistry, and the Center for Nanoscience and Nanotechnology at Tel-Aviv University. Work at Weizmann was supported by the Israel Science Foundation and the Lise-Meitner Minerva Center for Computational Quantum Chemistry.

## Appendix A – A Heuristic Derivation of the Driven Liouville von-Neumann Equation

The driven Liouville von-Neumann equation, presented in Eq. (4) of the main text, aims at modeling the full quantum system, consisting of semi-infinite leads and a molecule, which serves as the active device, using a closed (rather than an open) model system. The main approximation in such a description involves the replacement of the semi-infinite leads with finite lead models. In the full system, the semi-infinite leads play several important roles including: (i) altering the electronic properties of the active molecular entity; (ii) serving as electron reservoirs that absorb any incoming electron, destroy its phase, and prevent it from backscattering into the active molecular region; and (iii) inject incoherent electrons into the active molecular region with the appropriate Fermi-Dirac energy distribution according to the relevant chemical potential and electronic temperature. In order to provide a reliable description of the full infinite system, finite lead models must assume these roles in the closed system treatment. The first point mentioned above can be quite readily handled in the closed system description via the concept of the extended molecule where the active molecule is augmented by lead sections from its adjacent neighborhood. These lead sections are chosen to be sufficiently large such that the electronic properties of the resulting extended molecule are converged, to within the required accuracy, with respect to their size. Hence, two main challenges remain in the closed system description: (i) how to absorb and dephase electrons that are traveling towards the finite model edges before they are reflected back into the extended molecule region; and (ii) how to model the injection of incoherent electrons with the appropriate energy distribution from the lead models into the active molecular region. In the following, we discuss how both challenges are addressed in the driven Liouville von-Neumann equation presented in the main text.

## Electron Absorption

To describe electron absorption in the finite lead model systems, we invoke the concept of complex absorbing potentials (CAPs).[44, 45, 92-95] Here, the Hamiltonian of the finite lead models is augmented with an imaginary potential that, when traversed by an electron, induces a decay of the wavefunction. In the presence of such CAPs the Hamiltonian blocks of the lead models in the state representation (see main text) are represented by complex diagonal matrices. Here, the real part of each diagonal element is an Eigen-energy of the finite lead model Hamiltonian and its imaginary counterpart corresponds to the electron absorption/damping rate (or inverse lifetime) at this energy, which mimics the imaginary part of the self-energy of the semi-infinite lead.[96] When considering a two-lead junction, the Hamiltonian matrix in the state representation of the finite model system with CAPs reads:

$$\widehat{\widetilde{H}} = \begin{pmatrix} \widehat{\widetilde{H}}_L - i\widehat{\widetilde{\Gamma}}_L & \widehat{\widetilde{V}}_{L,EM} & \widehat{0} \\ \widehat{\widetilde{V}}_{EM,L} & \widehat{\widetilde{H}}_{EM} & \widehat{\widetilde{V}}_{EM,R} \\ \widehat{0} & \widehat{\widetilde{V}}_{R,EM} & \widehat{\widetilde{H}}_R - i\widehat{\widetilde{\Gamma}}_R \end{pmatrix}, \tag{A1}$$

where $\widehat{\widetilde{H}}_{L/R}$ are the diagonal matrices representing the discrete Eigen-spectrum of the finite left/right lead models, $\widehat{\widetilde{H}}_{EM}$ is a diagonal matrix representing the eigenvalues of the extended molecule, $\widehat{\widetilde{V}}_{L/R,EM}$ are the corresponding coupling matrices between the energy manifolds of the left/right leads and the extended molecule, $\widehat{\widetilde{V}}_{EM,L/R} = \left[\widehat{\widetilde{V}}_{L/R,EM}\right]^\dagger$, $\widehat{\widetilde{\Gamma}}_{L/R}$ are the absorbing potential diagonal matrix representations, and we assume that there is no direct coupling between the two leads.

Note that as in the main text, the tilde signs are used to designate matrices presented in the state-representation. In the wide-band limit one assumes that the electron absorption rate is independent of energy such that the $\hat{\tilde{\Gamma}}$ matrices are given by:

$$\hat{\tilde{\Gamma}}_{L/R} = \gamma_{L/R}\hat{I}_{L/R}, \tag{A2}$$

with $\gamma_{L/R}$ being the constant damping rate of the left/right leads and $\hat{I}_{L/R}$ are unit matrices with the dimensions of the corresponding left/right blocks.

The Hamiltonian of the full closed system can be now divided into its real and imaginary parts as follows:

$$\hat{\tilde{H}} = \hat{\tilde{H}}^{Re} - i\hat{\tilde{H}}^{Im} = \begin{pmatrix} \hat{\tilde{H}}_L & \hat{\tilde{V}}_{L,EM} & \hat{0} \\ \hat{\tilde{V}}_{EM,L} & \hat{\tilde{H}}_{EM} & \hat{\tilde{V}}_{EM,R} \\ \hat{0} & \hat{\tilde{V}}_{R,EM} & \hat{\tilde{H}}_R \end{pmatrix} - i\begin{pmatrix} \gamma_L\hat{I}_L & \hat{0} & \hat{0} \\ \hat{0} & \hat{0} & \hat{0} \\ \hat{0} & \hat{0} & \gamma_R\hat{I}_R \end{pmatrix}. \tag{A3}$$

Before studying the dynamics of the reduced density matrix under this Hamiltonian structure we must rewrite the Liouville von-Neumann equation for complex Hamiltonian matrix representations. To this end, we recall that the density operator is defined as

$$\hat{\rho}(t) = |\Psi(t)\rangle\langle\Psi(t)|, \tag{A4}$$

and hence its time derivative is given by:

$$\frac{d\hat{\rho}(t)}{dt} = \frac{d|\Psi(t)\rangle}{dt}\langle\Psi(t)| + |\Psi(t)\rangle\frac{d\langle\Psi(t)|}{dt}. \tag{A5}$$

The dynamics of the wave-function is dictated by the time-dependent Schrodinger equation, which for the ket state reads:

$$\frac{d|\Psi(t)\rangle}{dt} = -\frac{i}{\hbar}\hat{H}|\Psi(t)\rangle, \tag{A6}$$

and for the corresponding bra state is given by

$$\frac{d\langle\Psi(t)|}{dt} = \frac{i}{\hbar}\langle\Psi(t)|\hat{H}^\dagger. \tag{A7}$$

Using Eqs. (A6) and (A7) we can rewrite Eq. (A5) as:

$$\frac{d\hat{\rho}(t)}{dt} = -\frac{i}{\hbar}\hat{H}|\Psi(t)\rangle\langle\Psi(t)| + \frac{i}{\hbar}|\Psi(t)\rangle\langle\Psi(t)|\hat{H}^\dagger = -\frac{i}{\hbar}\left[\hat{H}\hat{\rho}(t) - \hat{\rho}(t)\hat{H}^\dagger\right] \tag{A8}$$

Because the real part of the Hamiltonian matrix is Hermitian and, in the state representation, the imaginary part is diagonal we obtain that $\hat{\tilde{H}}^\dagger = \hat{\tilde{H}}^{Re} + i\hat{\tilde{H}}^{Im}$. Hence, the time evolution of the density operator in this representation is given by:

$$\frac{d\hat{\tilde{\rho}}(t)}{dt} = -\frac{i}{\hbar}\left(\hat{\tilde{H}}^{Re} - i\hat{\tilde{H}}^{Im}\right)\hat{\tilde{\rho}}(t) + \frac{i}{\hbar}\hat{\tilde{\rho}}(t)\left(\hat{\tilde{H}}^{Re} + i\hat{\tilde{H}}^{Im}\right) =$$
$$= \left[-\frac{i}{\hbar}\hat{\tilde{H}}^{Re}\hat{\tilde{\rho}}(t) + \frac{i}{\hbar}\hat{\tilde{\rho}}(t)\hat{\tilde{H}}^{Re}\right] - \frac{1}{\hbar}\left[\hat{\tilde{H}}^{Im}\hat{\tilde{\rho}}(t) + \hat{\tilde{\rho}}(t)\hat{\tilde{H}}^{Im}\right] = -\frac{i}{\hbar}\left[\hat{\tilde{H}}^{Re}, \hat{\tilde{\rho}}(t)\right]_{-} - \frac{1}{\hbar}\left[\hat{\tilde{H}}^{Im}, \hat{\tilde{\rho}}(t)\right]_{+}$$
(A9)

where $[\ ]_{-}$ stands for the commutator and $[\ ]_{+}$ is the anti-commutator. The first term in the right hand side of Eq. (A9) represents the standard Liouville dynamics for the closed system, whereas the second term induces electron absorption at the finite lead models. Using the structure of $\hat{\tilde{H}}^{Im}$ presented in Eq. (A3) we can now evaluate the anti-commutator as follows:

$$-\frac{1}{\hbar}\left[\hat{\tilde{H}}^{Im}, \hat{\tilde{\rho}}(t)\right]_{+} = -\frac{1}{\hbar}\left[\hat{\tilde{H}}^{Im}\hat{\tilde{\rho}}(t) + \hat{\tilde{\rho}}(t)\hat{\tilde{H}}^{Im}\right] =$$
$$= -\frac{1}{\hbar}\begin{pmatrix} \gamma_L \hat{I}_L & \hat{0} & \hat{0} \\ \hat{0} & \hat{0} & \hat{0} \\ \hat{0} & \hat{0} & \gamma_R \hat{I}_R \end{pmatrix}\begin{pmatrix} \hat{\tilde{\rho}}_L & \hat{\tilde{\rho}}_{L,EM} & \hat{\tilde{\rho}}_{L,R} \\ \hat{\tilde{\rho}}_{EM,L} & \hat{\tilde{\rho}}_{EM} & \hat{\tilde{\rho}}_{EM,R} \\ \hat{\tilde{\rho}}_{R,L} & \hat{\tilde{\rho}}_{R,EM} & \hat{\tilde{\rho}}_R \end{pmatrix} - \frac{1}{\hbar}\begin{pmatrix} \hat{\tilde{\rho}}_L & \hat{\tilde{\rho}}_{L,EM} & \hat{\tilde{\rho}}_{L,R} \\ \hat{\tilde{\rho}}_{EM,L} & \hat{\tilde{\rho}}_{EM} & \hat{\tilde{\rho}}_{EM,R} \\ \hat{\tilde{\rho}}_{R,L} & \hat{\tilde{\rho}}_{R,EM} & \hat{\tilde{\rho}}_R \end{pmatrix}\begin{pmatrix} \gamma_L \hat{I}_L & \hat{0} & \hat{0} \\ \hat{0} & \hat{0} & \hat{0} \\ \hat{0} & \hat{0} & \gamma_R \hat{I}_R \end{pmatrix} =$$ (A10)
$$= -\frac{1}{\hbar}\begin{pmatrix} 2\gamma_L \hat{\tilde{\rho}}_L & \gamma_L \hat{\tilde{\rho}}_{L,EM} & (\gamma_L + \gamma_R)\hat{\tilde{\rho}}_{L,R} \\ \gamma_L \hat{\tilde{\rho}}_{EM,L} & \hat{0} & \gamma_R \hat{\tilde{\rho}}_{EM,R} \\ (\gamma_R + \gamma_L)\hat{\tilde{\rho}}_{R,L} & \gamma_R \hat{\tilde{\rho}}_{R,EM} & 2\gamma_R \hat{\tilde{\rho}}_R \end{pmatrix}$$

If we now choose $\gamma_L = \gamma_R = \gamma = \frac{1}{2}\Gamma$, where $\Gamma = \gamma_L + \gamma_R$ we obtain from Eq. (A9):

$$-\frac{1}{\hbar}\left[\hat{\tilde{H}}^{Im}, \hat{\tilde{\rho}}(t)\right]_{+} = -\frac{1}{\hbar}\Gamma\begin{pmatrix} \hat{\tilde{\rho}}_L & \frac{1}{2}\hat{\tilde{\rho}}_{L,EM} & \hat{\tilde{\rho}}_{L,R} \\ \frac{1}{2}\hat{\tilde{\rho}}_{EM,L} & \hat{0} & \frac{1}{2}\hat{\tilde{\rho}}_{EM,R} \\ \hat{\tilde{\rho}}_{R,L} & \frac{1}{2}\hat{\tilde{\rho}}_{R,EM} & \hat{\tilde{\rho}}_R \end{pmatrix}$$ (A11)

### **Electron Injection**

In the above derivation we have considered the absorption of electrons by the finite lead models. These act as the real (semi-)infinite lead electronic reservoirs by preventing electrons that enter the lead region from backscattering into the extended molecule. As mentioned earlier, the lead models must also inject electrons into the extended molecule region with the appropriate equilibrium Fermi-Dirac energy distribution of the real semi-infinite lead reservoirs, encoding their respective chemical potentials and electronic temperatures.

In the real system, the electrons are transferred from the semi-infinite leads into the extended molecule region without affecting the equilibrium state of the former. If one formally divides the full system into the semi-infinite leads regions and the extended molecule, this transfer process may be viewed as electron absorption at the lead surface interfacing the extended molecule and electron injection at the corresponding surface of the extended molecule region. Since the absorption and injection rates at the two sides of this imaginary interface are of equal magnitude and opposite signs we may describe the injection of electrons into the extended molecule region by considering the absorption of electrons traveling from deep inside the lead towards the extended molecule region at the semi-infinite lead surface. Following Eq. (A9), electron absorption in the finite lead models is described by the anti-commutator $-\frac{1}{\hbar}\left[\hat{\tilde{H}}^{\text{Im}},\hat{\tilde{\rho}}^0_{L/R}\right]_+$, where the time-dependent density matrix appearing in Eq. (A9) is replaced by the equilibrium density matrix of the relevant lead model, reflecting the fact that the leads remain at equilibrium despite exchanging electrons with the extended molecule. For the isolated semi-infinite leads within the wide band approximation this anti-commutator reads:

$$-\frac{1}{\hbar}\left[\hat{\tilde{H}}^{\text{Im}},\hat{\tilde{\rho}}^0_{L/R}\right]_+ =$$

$$= -\frac{1}{\hbar}\begin{pmatrix}\gamma_L\hat{I}_L & \hat{0} & \hat{0} \\ \hat{0} & \hat{0} & \hat{0} \\ \hat{0} & \hat{0} & \gamma_R\hat{I}_R\end{pmatrix}\begin{pmatrix}\hat{\tilde{\rho}}^0_L & \hat{0} & \hat{0} \\ \hat{0} & \hat{0} & \hat{0} \\ \hat{0} & \hat{0} & \hat{\tilde{\rho}}^0_R\end{pmatrix} - \frac{1}{\hbar}\begin{pmatrix}\hat{\tilde{\rho}}^0_L & \hat{0} & \hat{0} \\ \hat{0} & \hat{0} & \hat{0} \\ \hat{0} & \hat{0} & \hat{\tilde{\rho}}^0_R\end{pmatrix}\begin{pmatrix}\gamma_L\hat{I}_L & \hat{0} & \hat{0} \\ \hat{0} & \hat{0} & \hat{0} \\ \hat{0} & \hat{0} & \gamma_R\hat{I}_R\end{pmatrix} = \qquad (A12)$$

$$= -\frac{1}{\hbar}\begin{pmatrix}2\gamma_L\hat{\tilde{\rho}}^0_L & \hat{0} & \hat{0} \\ \hat{0} & \hat{0} & \hat{0} \\ \hat{0} & \hat{0} & 2\gamma_R\hat{\tilde{\rho}}^0_R\end{pmatrix} = -\frac{1}{\hbar}\Gamma\begin{pmatrix}\hat{\tilde{\rho}}^0_L & \hat{0} & \hat{0} \\ \hat{0} & \hat{0} & \hat{0} \\ \hat{0} & \hat{0} & \hat{\tilde{\rho}}^0_R\end{pmatrix},$$

where we must take the electron injection rates ($\gamma_L$ and $\gamma_R$) to be identical to those used above for the electron absorption term in order to maintain electron balance at the system-lead interface. Hence, adding this term, with opposite sign, to Eq. (A9) introduces electron emission with the appropriate Fermi-Dirac distribution at the imaginary boundaries of the extended molecule.

**Form of the Driven Liouville von-Neumann Equation**

From Eqs. (A9), (A11), and (A12) we obtain the following equation for the two-lead setup:

$$\frac{d\hat{\tilde{\rho}}(t)}{dt} = -\frac{i}{\hbar}\left[\hat{\tilde{H}}^{\text{Re}},\hat{\tilde{\rho}}(t)\right]_- -\frac{1}{\hbar}\left[\hat{\tilde{H}}^{\text{Im}},\hat{\tilde{\rho}}(t)\right]_+ +\frac{1}{\hbar}\left[\hat{\tilde{H}}^{\text{Im}},\hat{\tilde{\rho}}^0_L\right]_+ +\frac{1}{\hbar}\left[\hat{\tilde{H}}^{\text{Im}},\hat{\tilde{\rho}}^0_R\right]_+ =$$

$$= -\frac{i}{\hbar}\left[\hat{\tilde{H}}^{\text{Re}},\hat{\tilde{\rho}}(t)\right]_- -\frac{1}{\hbar}\Gamma\begin{pmatrix}\hat{\tilde{\rho}}_L - \hat{\tilde{\rho}}^0_L & \tfrac{1}{2}\hat{\tilde{\rho}}_{L,EM} & \hat{\tilde{\rho}}_{L,R} \\ \tfrac{1}{2}\hat{\tilde{\rho}}_{EM,L} & \hat{0} & \tfrac{1}{2}\hat{\tilde{\rho}}_{EM,R} \\ \hat{\tilde{\rho}}_{R,L} & \tfrac{1}{2}\hat{\tilde{\rho}}_{R,EM} & \hat{\tilde{\rho}}_R - \hat{\tilde{\rho}}^0_R\end{pmatrix}, \qquad (A13)$$

which is our working equation in the state representation, namely Eq. (9) of the main text. This equation can be readily transformed to the atomistic/site-representation using the global unitary transformation matrix of Eq. (6). Following the definition $\hat{\tilde{\rho}} = \hat{U}^\dagger \hat{\rho} \hat{U}$, appearing in Eq. (8) of the main text, and the fact that $\hat{U}^\dagger \hat{U} = \hat{U}\hat{U}^\dagger = \hat{I}$, we may write $\hat{U}\hat{\tilde{\rho}}\hat{U}^\dagger = \hat{U}\hat{U}^\dagger \hat{\rho} \hat{U} \hat{U}^\dagger = \hat{\rho}$. Hence,

$$\frac{d\hat{\rho}(t)}{dt} = \frac{d\hat{U}\hat{\tilde{\rho}}(t)\hat{U}^\dagger}{dt} = \hat{U}\frac{d\hat{\tilde{\rho}}(t)}{dt}\hat{U}^\dagger = -\frac{i}{\hbar}\hat{U}\left[\hat{\tilde{H}}^{\text{Re}},\hat{\tilde{\rho}}(t)\right]_-\hat{U}^\dagger - \frac{1}{\hbar}\Gamma\hat{U}\begin{pmatrix}\hat{\tilde{\rho}}_L - \hat{\tilde{\rho}}_L^0 & \frac{1}{2}\hat{\tilde{\rho}}_{L,EM} & \hat{\tilde{\rho}}_{L,R} \\ \frac{1}{2}\hat{\tilde{\rho}}_{EM,L} & \hat{0} & \frac{1}{2}\hat{\tilde{\rho}}_{EM,R} \\ \hat{\tilde{\rho}}_{R,L} & \frac{1}{2}\hat{\tilde{\rho}}_{R,EM} & \hat{\tilde{\rho}}_R - \hat{\tilde{\rho}}_R^0\end{pmatrix}\hat{U}^\dagger =$$

$$= -\frac{i}{\hbar}\hat{U}\hat{\tilde{H}}^{\text{Re}}\hat{\tilde{\rho}}(t)\hat{U}^\dagger + \frac{i}{\hbar}\hat{U}\hat{\tilde{\rho}}(t)\hat{\tilde{H}}^{\text{Re}}\hat{U}^\dagger - \frac{1}{\hbar}\Gamma\begin{pmatrix}\hat{U}_L & \hat{0} & \hat{0} \\ \hat{0} & \hat{U}_{EM} & \hat{0} \\ \hat{0} & \hat{0} & \hat{U}_R\end{pmatrix}\begin{pmatrix}\hat{\tilde{\rho}}_L - \hat{\tilde{\rho}}_L^0 & \frac{1}{2}\hat{\tilde{\rho}}_{L,EM} & \hat{\tilde{\rho}}_{L,R} \\ \frac{1}{2}\hat{\tilde{\rho}}_{EM,L} & \hat{0} & \frac{1}{2}\hat{\tilde{\rho}}_{EM,R} \\ \hat{\tilde{\rho}}_{R,L} & \frac{1}{2}\hat{\tilde{\rho}}_{R,EM} & \hat{\tilde{\rho}}_R - \hat{\tilde{\rho}}_R^0\end{pmatrix}\begin{pmatrix}\hat{U}_L^\dagger & \hat{0} & \hat{0} \\ \hat{0} & \hat{U}_{EM}^\dagger & \hat{0} \\ \hat{0} & \hat{0} & \hat{U}_R^\dagger\end{pmatrix} =$$

$$= -\frac{i}{\hbar}\hat{U}\hat{\tilde{H}}^{\text{Re}}\hat{U}^\dagger\hat{U}\hat{\tilde{\rho}}(t)\hat{U}^\dagger + \frac{i}{\hbar}\hat{U}\hat{\tilde{\rho}}(t)\hat{U}^\dagger\hat{U}\hat{\tilde{H}}^{\text{Re}}\hat{U}^\dagger - \frac{1}{\hbar}\Gamma\begin{pmatrix}\hat{U}_L(\hat{\tilde{\rho}}_L - \hat{\tilde{\rho}}_L^0)\hat{U}_L^\dagger & \frac{1}{2}\hat{U}_L\hat{\tilde{\rho}}_{L,EM}\hat{U}_{EM}^\dagger & \hat{U}_L\hat{\tilde{\rho}}_{L,R}\hat{U}_R^\dagger \\ \frac{1}{2}\hat{U}_{EM}\hat{\tilde{\rho}}_{EM,L}\hat{U}_L^\dagger & \hat{0} & \frac{1}{2}\hat{U}_{EM}\hat{\tilde{\rho}}_{EM,R}\hat{U}_R^\dagger \\ \hat{U}_R\hat{\tilde{\rho}}_{R,L}\hat{U}_L^\dagger & \frac{1}{2}\hat{U}_R\hat{\tilde{\rho}}_{R,EM}\hat{U}_{EM}^\dagger & \hat{U}_R(\hat{\tilde{\rho}}_R - \hat{\tilde{\rho}}_R^0)\hat{U}_R^\dagger\end{pmatrix} =$$

$$= -\frac{i}{\hbar}\hat{H}^{\text{Re}}\hat{\rho}(t) + \frac{i}{\hbar}\hat{\rho}(t)\hat{H}^{\text{Re}} - \frac{1}{\hbar}\Gamma\begin{pmatrix}\hat{\rho}_L - \hat{\rho}_L^0 & \frac{1}{2}\hat{\rho}_{L,EM} & \hat{\rho}_{L,R} \\ \frac{1}{2}\hat{\rho}_{EM,L} & \hat{0} & \frac{1}{2}\hat{\rho}_{EM,R} \\ \hat{\rho}_{R,L} & \frac{1}{2}\hat{\rho}_{R,EM} & \hat{\rho}_R - \hat{\rho}_R^0\end{pmatrix} =$$

$$= -\frac{i}{\hbar}\left[\hat{H}^{\text{Re}},\hat{\rho}(t)\right]_- - \frac{1}{\hbar}\Gamma\begin{pmatrix}\hat{\rho}_L - \hat{\rho}_L^0 & \frac{1}{2}\hat{\rho}_{L,EM} & \hat{\rho}_{L,R} \\ \frac{1}{2}\hat{\rho}_{EM,L} & \hat{0} & \frac{1}{2}\hat{\rho}_{EM,R} \\ \hat{\rho}_{R,L} & \frac{1}{2}\hat{\rho}_{R,EM} & \hat{\rho}_R - \hat{\rho}_R^0\end{pmatrix} \qquad (A14)$$

where we have used the fact that the transformation matrix, $\hat{U}$, is time-independent, and that the individual blocks of the density matrix transform as $\hat{\tilde{\rho}}_{i,j} = \hat{U}_i^\dagger \hat{\rho}_{i,j} \hat{U}_j$ such that $\hat{U}_i \hat{\tilde{\rho}}_{i,j} \hat{U}_j^\dagger = \hat{U}_i \hat{U}_i^\dagger \hat{\rho}_{i,j} \hat{U}_j \hat{U}_j^\dagger = \hat{\rho}_{i,j}$. Eq. (A14) is, in fact, Eq. (4) of the main text.

We note that in the present derivation we used the concept of complex absorbing potentials and worked within the wide band approximation in order to obtain a simple form of our working equation, involving a single driving rate. Nevertheless, given the exact form of the lead self-energies, one can avoid the approximations involved with the choice of CAPs and repeat the derivation with the explicit energy dependent absorbing terms going beyond the wide band limit and obtaining a more general equation. Here, however, one will have to explicitly calculate the energy dependent lead self-energies for each choice of lead model.

## **Appendix B – Calculating Bond Currents in Tight-Binding Models**

In the main text we have used the off-diagonal elements of the density matrix to calculate the bond currents in the various model systems considered. The relevant formula can be derived from the fact that in a tight-binding model the molecular orbitals ($\psi_i$) are represented as a linear combination of atomic orbitals ($\varphi_n$) in the following manner:

$$|\psi_i(t)\rangle = \sum_n C_{ni}(t)|\varphi_n\rangle. \qquad (B1)$$

Because we are considering non-interacting electron models, the overall current is obtained from the sum of the contribution of the individual electrons or the individual occupied molecular orbitals. The time dependence of each orbital is given by the single electron time-dependent Schrödinger equation:

$$\frac{d|\psi_i(t)\rangle}{dt} = -\frac{i}{\hbar}\hat{H}|\psi_i(t)\rangle. \tag{B2}$$

Inserting the basis set expansion we obtain:

$$\sum_n \frac{dC_{ni}(t)}{dt}|\varphi_n\rangle = -\frac{i}{\hbar}\sum_n C_{ni}(t)\hat{H}|\varphi_n\rangle. \tag{B3}$$

Now, operating with $\langle\varphi_m|$ from the left one obtains:

$$\sum_n \frac{dC_{ni}(t)}{dt}\langle\varphi_m|\varphi_n\rangle = -\frac{i}{\hbar}\sum_n C_{ni}(t)\langle\varphi_m|\hat{H}|\varphi_n\rangle. \tag{B4}$$

In the tight-binding model the basis is orthonormal and the Hamiltonian matrix is tridiagonal, i.e.,

$$\langle\varphi_m|\varphi_n\rangle = \delta_{m,n} \tag{B5}$$

and:

$$\langle\varphi_m|\hat{H}|\varphi_n\rangle = H_{n,n}\delta_{m,n} + H_{n-1,n}\delta_{m,n-1} + H_{n+1,n}\delta_{m,n+1}. \tag{B6}$$

We therefore obtain:

$$\frac{dC_{mi}(t)}{dt} = -\frac{i}{\hbar}\left[C_{mi}(t)H_{m,m} + C_{m+1,i}(t)H_{m,m+1} + C_{m-1,i}(t)H_{m,m-1}\right]. \tag{B7}$$

By conjugating this equation one gets:

$$\frac{dC^*_{mi}(t)}{dt} = +\frac{i}{\hbar}\left[C^*_{mi}(t)H_{m,m} + C^*_{m+1,i}(t)H^*_{m,m+1} + C^*_{m-1,i}(t)H^*_{m,m-1}\right]. \tag{B8}$$

The population dynamics on each site can now be readily calculated. At a given time, t, the contribution of a given orbital $\psi_i(t)$ to the occupation of site $n$ in the tight-binding chain is given by $|C_{ni}(t)|^2$. From the continuity equation the time derivative of this occupation is the difference between the incoming and outgoing currents. We therefore calculate the time derivative of the single orbital contribution as:

$$\frac{d|C_{ni}(t)|^2}{dt} = \frac{d[C^*_{ni}(t)C_{ni}(t)]}{dt} = C^*_{ni}(t)\frac{d[C_{ni}(t)]}{dt} + C_{ni}(t)\frac{d[C^*_{ni}(t)]}{dt}. \tag{B9}$$

Using Eqs. (B7) and (B8) in (B9) we arrive at:

$$\frac{d|C_{ni}(t)|^2}{dt} =$$

$$= \frac{i}{\hbar}\left[C_{ni}(t)C^*_{n+1,i}(t)H^*_{n,n+1} - C^*_{ni}(t)C_{n+1,i}(t)H_{n,n+1}\right] + \frac{i}{\hbar}\left[C_{ni}(t)C^*_{n-1,i}(t)H^*_{n,n-1} - C^*_{ni}(t)C_{n-1,i}(t)H_{n,n-1}\right]. \tag{B10}$$

Assuming that all non-zero off-diagonal terms are real, identical, and with a value of $\beta$ we get:

$$\frac{d|C_{ni}(t)|^2}{dt} = \frac{i\beta}{\hbar}\{2i * Im[C_{ni}(t)C^*_{n+1,i}(t)] + 2i * Im[C_{ni}(t)C^*_{n-1,i}(t)]\} =$$
$$-\frac{2\beta}{\hbar}[Im[C_{ni}(t)C^*_{n+1,i}(t)] + Im[C_{ni}(t)C^*_{n-1,i}(t)]] = -\frac{2\beta}{\hbar}[Im[C_{ni}(t)C^*_{n+1,i}(t)] - Im[C_{n-1,i}(t)C^*_{ni}(t)]] \quad (B11)$$

In analogy to the continuity equation $\left(\frac{d\rho}{dt} = -\vec{\nabla}\cdot\vec{j}\right)$ we can now identify $\frac{2\beta}{\hbar}Im[C_{ni}(t)C^*_{n+1,i}(t)]$ as the particle current going from site n to site n+1 or the n,n+1 bond particle current and $\frac{2\beta}{\hbar}Im[C_{n-1,i}(t)C^*_{ni}(t)]$ as the particle current going from site n-1 to site n or the n-1,n bond particle current. Multiplying by the electron charge $|e|$ to obtain the electrical current from the particle current, summing over the contribution of all occupied molecular orbitals, and using the density matrix notation we obtain the following expression for the bond current:

$$J_{n,n+1} = \frac{2\beta|e|}{\hbar}\cdot 2\times\sum_{i=1}^{N_{occ}} Im[C_{ni}(t)C^*_{n+1,i}(t)] = \frac{2\beta|e|}{\hbar}Im[\rho_{n,n+1}], \quad (B12)$$

where we have assumed a closed shell system and the factor of 2 before the summation stands for the contribution of 2 electrons occupying each occupied molecular orbital.

## **Appendix C – Description of the Landauer Transport Calculations**

To perform the steady-state reference calculations we followed the standard Landauer formalism combined with non-equilibrium Green's functions theory to calculate the corresponding transmittance probabilities, with some modifications in the construction of the leads' Green's functions, as detailed below. Here the current (I) flowing between the left ($L$) and the right ($R$) leads at a given bias voltage (V) is calculated via the probability ($T$) of an electron with a given energy ($E$) to traverse the system:

$$I(V) = \frac{2e}{h}\int_{-\infty}^{\infty}dE[f_L(E,\mu_L(V)) - f_R(E,\mu_R(V))]T(E). \quad (C1)$$

Here, e is the electron charge, h is Plank's constant, $f_{L/R}$ are the Fermi-Dirac distributions functions representing the electron occupations in the left/right leads given by $f_{L/R}(E,\mu_{L/R}(V)) = \left[1+e^{\beta_{L/R}(E-\mu_{L/R}(V))}\right]$, $\beta_{L/R} = \frac{1}{k_B T_{L/R}}$, $k_B$ is Boltzmann's constant, $T_{L/R}$ - the electronic temperature in the left/right leads, and the bias voltage is assumed to drop symmetrically at the lead-molecule junctions such that the leads' chemical potentials $\mu_{L/R}$ are chosen as $\mu_{L/R} = E_f^{L/R} \pm 0.5V$, with $E_f^{L/R}$ being the Fermi energy of the left/right lead. The transmission probability is calculated via the following trace formula, which can be derived using non-equilibrium Green's functions techniques:[16, 17, 97]

$$T(E) = Tr[\hat{\Gamma}_L(E)\hat{G}^r_{EM}(E)\hat{\Gamma}_R(E)\hat{G}^a_{EM}(E)]. \quad (C2)$$

Here, $\widehat{G}_{EM}^r(E) = [E\hat{I} - \widehat{H}_{EM} - \Sigma_L^r(E) - \Sigma_R^r(E)]^{-1}$ is the retarded Green's function of the extended molecule, where $\widehat{H}_{EM}$ is the Hamiltonian matrix representation of the extended molecule, $\hat{I}$ is a unit matrix of the same dimensions, and the lead's self-energy functions $\Sigma_{L/R}^r(E)$ are given by:

$$\widehat{\Sigma}_{L/R}^r(E) = (E\hat{I} - \widehat{V}_{EM,L/R})\widehat{G}_{L/R}^r(E)(E\hat{I} - \widehat{V}_{L/R,EM}). \tag{C3}$$

$\widehat{V}_{EM,L/R}$ is the Hamiltonian matrix block representing the coupling between the extended molecule and the left/right lead, $\widehat{V}_{L/R,EM} = [\widehat{V}_{EM,L/R}]^\dagger$, and $\widehat{G}_{L/R}^r(E)$ is the retarded surface Green's function of the bare semi-infinite lead. The advanced Green's function matrix representation of the extended molecule is given by $\widehat{G}_{EM}^a(E) = [\widehat{G}_{EM}^r(E)]^\dagger$ and the broadening functions, $\widehat{\Gamma}_{L/R}(E)$, are given by:

$$\widehat{\Gamma}_{L/R}(E) = i[\widehat{\Sigma}_{L/R}^r(E) - \widehat{\Sigma}_{L/R}^a(E)], \tag{C4}$$

with $\widehat{\Sigma}_{L/R}^a(E) = [\widehat{\Sigma}_{L/R}^r(E)]^\dagger$.

The retarded surface Green's function of the bare semi-infinite lead, $\widehat{G}_{L/R}^r(E)$, can be solved for by using efficient iterative methods.[98-103] Here, in the spirit of the finite model system, we calculate it by complex matrix inversion of $\widehat{G}_{L/R}^r(E) = [E\hat{I} - \widehat{H}_{L/R} + i\eta]^{-1}$, where $\widehat{H}_{L/R}$ is the Hamiltonian matrix representation of the finite left/right lead model and $i\eta$ is a small imaginary broadening factor introduced to eliminate the singularities and serves to broaden the discrete spectra of the finite lead models into a quasi-continuous one. To this end, $\eta$ is chosen as twice the maximum eigenenergy spacing within the corresponding lead model. The obtained Landauer current is then converged with respect to the size of the lead models to the required accuracy.

In a multi-lead setup the current flowing between any two leads can be calculated in the same manner presented above for the two-lead setup, where the retarded Green's function of the extended molecule has to include the self-energy contributions of all the leads:

$$\widehat{G}_{EM}^r(E) = [E\hat{I} - \widehat{H}_{EM} - \sum_{i=1}^{N_{leads}} \Sigma_i^r(E)]^{-1} \tag{C5}$$

We note that, owing to the tight-binding nature of our model, the Landauer calculations we perform are not self-consistent in terms of the influence of the bias voltage on the eigenstates of the system. This is consistent with the fact that in the dynamical calculations the Hamiltonian is taken to be time-independent as well. Naturally, if one uses more complex schemes, such as time-dependent density functional theory, to perform the dynamics, the corresponding steady-state Landauer calculations must be done self-consistently.